\documentclass[12pt]{article}
\usepackage{amsmath}
\usepackage{times}
\usepackage{graphicx}
\usepackage{color}
\usepackage{multirow}
\usepackage[authoryear]{natbib}
\usepackage{rotating}
\usepackage{bbm}
\usepackage{latexsym}
\newcommand{\captionfonts}{\normalsize}

\makeatletter  
\long\def\@makecaption#1#2{%
  \vskip\abovecaptionskip
  \sbox\@tempboxa{{\captionfonts #1: #2}}%
  \ifdim \wd\@tempboxa >\hsize
    {\captionfonts #1: #2\par}
  \else
    \hbox to\hsize{\hfil\box\@tempboxa\hfil}%
  \fi
  \vskip\belowcaptionskip}
\makeatother   
\begin{document}
\hspace{13.9cm}1

\ \vspace{20mm}\\

%Dual Dimension Reduction is more Data Efficient than Independent Dimensionality Reduction: A Lesson From the Symmetric Information Bottleneck
{\LARGE \noindent Data efficiency, dimensionality reduction, and the generalized symmetric information bottleneck
}

\ \\
{\bf \large K. Michael Martini$^{\displaystyle 1, \displaystyle 3}$ 
 and Ilya Nemenman$^{\displaystyle 1,\displaystyle 2, \displaystyle 3}$}\\
{$^{\displaystyle 1}$ Department of Physics, Emory University, Atlanta, GA 30322.}\\
{$^{\displaystyle 2}$ Department of Biology, Emory University, Atlanta, GA 30322.}\\
{$^{\displaystyle 3}$ Initiative in Theory and Modeling of Living Systems, Emory University, Atlanta, GA 30322.}\\
%

%\ \\[-2mm]
{\bf \noindent Keywords:} Information Bottleneck, Symmetric Information Bottleneck, Dimensionality Reduction, Error Bounds, Data Efficiency

\thispagestyle{empty}
\markboth{}{NC instructions}
\ \vspace{-0mm}\\
%
%Abstract
\begin{center} {\bf Abstract} \end{center}
The Symmetric Information Bottleneck (SIB), an extension of the more familiar Information Bottleneck, is a dimensionality reduction technique that simultaneously compresses two random variables to preserve information between their compressed versions.  We introduce the Generalized Symmetric Information Bottleneck (GSIB), which explores different functional forms of the cost of such simultaneous reduction. We then explore the dataset size requirements of such simultaneous compression.  We do this by deriving bounds and root-mean-squared estimates of statistical fluctuations of the involved loss functions. We show that, in typical situations, the simultaneous GSIB compression requires qualitatively less data to achieve the same errors compared to compressing variables one at a time.  We suggest that this is an example of a more general principle that simultaneous compression is more data efficient than independent compression of each of the input variables.

%%%%%%%%%%%

\section{Introduction}

Recent years have seen an explosion of large-dimensional experimental data sets \citep{de2020large,siegle2021survey,haghighi2022high} and the parallel growth in the number of methods for {\em dimensionality reduction} (DR)---that is, for extracting  low-dimensional structure from large-dimensional data \citep{carreira1997review, van2009dimensionality, nanga2021review}. Broadly speaking, we classify dimensionality reduction methods into two classes: unsupervised and supervised.  Unsupervised DR methods seek a low-dimensional description, $T_X$, of a large-dimensional variable, $X$, that preserves its variance, entropy, or another measure of diversity of the data. Such methods include the familiar principal component analysis (PCA) \citep{hotelling1933analysis}, non-negative matrix factorization \citep{lee1999learning}, multidimensional scaling (MDS) \citep{kruskal1964multidimensional}, t-distributed stochastic neighbor embedding (t-SNE) \citep{van2008visualizing}, Isomap \citep{tenenbaum2000global}, Uniform Manifold Approximation and Projection (UMAP) \citep{mcinnes2018umap}, autoencoders \citep{hinton2006reducing}, and related techniques \citep{Welling2014}. In contrast, supervised DR techniques aim to find a low-dimensional description, $T_X$, of a large dimensional $X$, while preserving $T_X$'s ability to explain another variable $Y$, which provides an effective {\em relevance} or {\em supervision} signal. Common examples include variable selection in regression \citep{andersen2010variable,kuo1998variable}, cross-encoders, Bayesian Ising Approximation (BIA) \citep{fisher2015bayesian}, and the Information Bottleneck (IB) \citep{tishby2000information, tishby2000data}. A particularly interesting class of such supervised dimensionality reduction problems is when both the reduced variable $X$ and the relevance variable $Y$ are large-dimensional. In these situations, finding significant correlations within combinatorially many groups of components of $X$ and $Y$ is hard, suggesting parallel dimensionality reduction of both $X$ and $Y$ into $T_X$ and $T_Y$, respectively.

We distinguish three classes of approaches to this problem. In the first, which we call the {\em Independent Unsupervised Dimensionality Reduction} (IUDR), one applies unsupervised DR methods to $X$ and $Y$ independently. One then searches for statistical dependencies between $T_X$ and $Y$ or $T_Y$ and $X$ or $T_X$ and $T_Y$, but the dimensionality reduction itself is agnostic of this subsequent step. A familiar example of this is the Principal Components Regression, where the projections on the principal components of $X$ are regressed against $Y$. We also distinguish {\em Independent Supervised Dimensionality Reduction} (ISDR), where $T_X$ is produced by compressing $X$ with $Y$ as the supervision signal, while $T_Y$ emerges from compressing $Y$ with $X$ as the supervision. The Information Bottleneck (IB) \citep{tishby2000information}, the Generalized and Deterministic Information Bottleneck (GIB) \citep{strouse2017deterministic}, and cross-encoders are examples of such approaches. Finally, {\em Simultaneous Supervised Dimensionality Reduction} (SSDR) is a class of methods where $T_X$ and $T_Y$ are produced simultaneously, typically being supervision signals of each other.\footnote{SSDR methods are sometimes referred to as {\em dual} DR \citep{sponberg2015dual}. We believe that the terminology we  propose here is better suited for classifying the breadth of DR approaches.} Examples of SSDR  include the Canonical Correlation Analysis (CCA) \citep{Hotelling1936, Tao2021} and its modern nonlinear neural network based generalizations \citep{Livescu2013, Wang2021}, Partial Least Squares (PLS) \citep{wold1966estimation, Eriksson2001}, and the Symmetric version of the Information Bottleneck (SIB) \citep{slonim2006multivariate}.\footnote{Different sources refer to CCA or PLS as both supervised or unsupervised techniques or something inbetween \citep{holbrook2017bayesian,scott2021using,zhuang2020technical}. Within our classification scheme, these are supervised methods.} In this paper we introduce a Generalized version of the Symmetric Information Bottleneck (GSIB) by interpolating between the compression cost measured by entropy and information.  This parallels for SSDR the introduction of the Generalized Information Bottleneck (GIB) for ISDR, of which the Deterministic Information Bottleneck and the Information Bottleneck are limits \citep{strouse2017deterministic}.

%more references for deepCCA: Livescu2016, Schuurmans2021,Tao2021
%more references for PLS \citep{Newsted1999, Eriksson2001,Thompson2006, Ye2009, Thompson2012}

We then argue that SSDR approaches can require a lot fewer data than their ISDR counterparts to achieve the same accuracy. We demonstrate this by comparing the bias and statistical fluctuations in the objective functions of independent GIB reductions of variables $X$ and $Y$ (ISDR approach) with the corresponding bias and fluctuations for the GSIB (SSDR approach). We show that the bias for the GSIB scales as the product of cardinalities of the compressed variables, while the bias for the GIB scales as the (typically much larger) product of cardinalities of the supervision signal and the compressed variable. We do the comparison for both typical fluctuations and for the upper bounds on the fluctuations. While our derivations are done for the IB approaches only, the intuitive explanation of the differences between the approaches suggests that SSDR methods are likely to require less data than their ISDR analogues more generally. 

\section{Background: Information Bottleneck and the Symmetric Information Bottleneck}

\subsection{Information Bottleneck and Its Generalizations} 
The goal of {\em Information Bottleneck} (IB) is to produce a compression, $T_X$ of a random variable $X$, such that the compression retains as much information as possible  about another  random variable $Y$, which is called the {\em relevant} (or, in our language, the {\em supervising}) variable. The information is measured using Shannon's mutual information \citep{shannon1948mathematical}, which quantifies the difference between the joint probability distribution $p(x,y)$ and the product of the marginal distributions $p(x)p(y)$:
\begin{equation}
I(X,Y)=\sum_{X,Y} p(x,y) \frac{\log(p(x,y))}{p(x)p(y)}=H(X)-H(X|Y),
\end{equation}
where $H(X)$ is the entropy of the variable $X$ and $H(X|Y)$ is the conditional entropy of $X$ given $Y$, $H(X|Y)=\sum_Y p(y) H(X|Y=y) = \sum_Y p(y) \sum_X p(x|y) \log(p(x|y))$. Mutual information is symmetric, always non-negative, and is  only zero when the random variables are independent \citep{cover1999elements}.

To achieve its goal, IB  produces a probabilistic mapping from $X$ to $T_X$, $p(t_x|x)$, which minimizes a specific cost function. The cost function  trades off preserving the information in the compression about the relevant variable,  $I(T_X,Y)$, against losing the information about $X$ (reducing the variable), $I(T_X,X)$:
\begin{equation}
    L_{\rm IB}=I(T_X,X)-\beta I(T_X,Y).
\end{equation}
Here $\beta$ is the trade-off parameter, which controls how important the compression $I(T_X,X)$ is compared to preserving the relevant information $I(T_X,Y)$. As $\beta\to\infty$, the  cost function is minimized by having no compression, $X=T$. Recently a Generalized version of IB was proposed (GIB) \citep{strouse2017deterministic},  which changes the cost function to
\begin{equation}
L_{\rm GIB}=H(T_X)-\alpha_x H(T_X|X)-\beta I(T_X,Y),
\end{equation}
which has a formal solution
\begin{align}
p(t_x|x)&=\frac{1}{Z(\beta,\alpha)}\exp\left[\frac{1}{\alpha_x}\left(\log p(t_x)-\beta D_{\rm KL}( p(y|x)||p(y|t_x))\right)\right],\\
  p(y|t_x)&=\frac{1}{p(t_x)}\sum_X p(t_x|x)p(x,y),\end{align}
where $D_{\rm KL}$ is the usual Kullback–Leibler divergence \citep{kullback1951information}.

The original IB is recovered from GIB  when $\alpha_x=1$. In contrast, when $\alpha_x\to0$,  $I(T_X,X)$ is replaced with $H(T_X)$ in the cost function. This  corresponds to replacing the cost of having a noisy channel encoding  $X$ into $T_X$ with the cost of directly storing  $T_X$. In this case, the formal solution results in a deterministic mapping between $X$ and $T_X$, and the resulting problem is known as the {\em Deterministic Information Bottleneck} (DIB) \citep{strouse2017deterministic}.

If both $X$ and $Y$ are large-dimensional and require dimensionality reduction, one can apply IB to produce the mapping $X\to T_X$ with $Y$ as the relevant variable, and then solve a separate IB problem to map $Y\to T_Y$ with $X$ as the supervision. This approach would fall into the ISDR class in our nomenclature.

\subsection{Symmetric Information Bottleneck and its Generalization}
The Symmetric Information Bottleneck (SIB), introduced in \cite{slonim2006multivariate}, is an SSDR approach, where $X$ and $Y$ are compressed simultaneously, such that the compressed versions $T_X$, and $T_Y$ contain the maximal amount of information about each other. This corresponds to optimizing the loss function:
\begin{equation}
L_{\rm SIB}=I(T_X;X)+I(T_Y;Y)-\beta I(T_X;T_Y),
\end{equation}
where optimization is over all possible probabilistic compressions $p(t_x|x)$ and $p(t_y|y)$. As before,  $\beta$ determines the strength of the trade-off between the compression and preserving the relevant information. 

For generality, here we propose a Generalized SIB (GSIB), which incorporates flexible compression terms, similar to how GIB was optained from IB. The new cost function is
\begin{align}
L_{\rm GSIB}&=I_{\alpha_X}(T_X;X)+I_{\alpha_Y}(T_Y;Y)-\beta I(T_X;T_Y)\\
&=H(T_X)-\alpha_X H(T_X|X) + H(T_Y)-\alpha_Y H(T_Y|Y) - \beta I(T_X, T_Y).
\label{eq:gsib}
\end{align}
Here we defined shorthands $I_{\alpha_X}(T_X,X)=H(T_X)-\alpha_X H(T_X|X)$, and similarly for $I_{\alpha_Y}$, and the  cost function must be minimized with  respect to $p(t_x|x)$ and $p(t_y|y)$. The parameters $\alpha_X$ and $\alpha_Y$ are what dictates how probabilistic the mapping between the uncompressed variables and their compressed versions is. In the limit $\alpha_x,\alpha_Y\to 0$, the mapping  can be verified to be deterministic (see below), resulting in the Determistic SIB (DSIB).  When $\alpha_X,\alpha_Y\to 1$, GSIB becomes the usual SIB.

Optimization of the cost function has a formal solution: 
\begin{align}
p(t_x|x)&=\frac{\exp\left[\frac{1}{\alpha_X}\left(\ln p(t_x)-\beta D_{\rm KL}(p(t_y|x)||p(t_y|t_x)\right)\right]}{Z_x(x,\alpha_X,\beta)},\\
p(t_y|y)&=\frac{\exp\left[\frac{1}{\alpha_Y}\left(\ln p(t_y)-\beta D_{\rm KL}(p(t_x|y)||p(t_x|t_y)\right)\right]}{Z_y(y,\alpha_Y,\beta)},\\
p(t_y|x)&=\frac{\sum_{Y}p(t_y|y)p(x,y)}{p(x)},\;\;\;\;\;
p(t_y|t_x)=\frac{\sum_{X,Y}p(t_y|y)p(t_x|x)p(x,y)}{\sum_{X} p(t_x|x)p(x)},\\
p(t_x|y)&=\frac{\sum_{X}p(t_x|x)p(x,y)}{p(y)},\;\;\;\;\;
p(t_x|t_y)=\frac{\sum_{X,Y}p(t_y|y)p(t_x|x)p(x,y)}{\sum_{Y} p(t_y|y)p(y)}.\\
\nonumber\end{align}
Similar to  IB, this formal solution can be iterated starting from an initial guess for both $p(t_x|x)$ and $p(t_y|y)$.

Interestingly, parenthetically we note that, unlike for IB, there are now  exponentially many, $\sim 2^{|T_X|+|T_Y|}$, trivial fixed points for this iteration scheme (here $|\cdot|$ denotes cardinality of the variable, so that the rest of our discussion focuses on random variables defined on discrete, finite sets of possible values). For example, a uniform distribution for both random mappings, $p(t_x|x)=1/|T_X|$ and $p(t_y|y)=1/|T_Y|$ is a fixed point of the iteration with the cost of zero, even though a uniform mapping, independent of the conditioning variable, is clearly not a useful compression. Furthermore, all  distributions, where $p(t_x|x)$ is zero for several values of $t_x$ and uniform otherwise, are also trivial fixed points. There are exponentially many distributions of this type. When $\alpha_x=\alpha_y=1$, these distributions are part of a larger class of trivial fixed points, which includes all mappings independent of the data, i.~e., $p(t_x|x)=A(t_x)$ and $p(t_y|y)=B(t_y)$. One can easily verify that the first derivative of $L_{\rm GSIB}$ vanishes for these solutions. The second derivative, which controls if these solutions are minima or maxima, is:
\begin{equation}\frac{\partial^2 L_{\rm GSIB}}{\partial  p(t_x|x) \partial  p(t_x'|x')}=\frac{-p(x)}{A(t_x)}(p(x)-\alpha_X)\delta(x,x')\delta(t,t_x')-\frac{p(x)p(x')}{A(t_x)}\delta(t_x,t_x')(1-\delta(x,x')),\label{secondOrder}\end{equation}
(with similar expression for the compression of $Y$).  These trivial fixed points are maxima when $\alpha_x<p(x)$, and $\alpha_y<p(y)$. When $\alpha_x>p(x)$ and $\alpha_y>p(y)$, such as in the case of SIB, when $\alpha_X=\alpha_Y=1$, the trivial fixed points are saddles. Thus solutions found by the iterative algorithm must be viewed with suspicion, and one should always verify if the algorithm got trapped by one of the trivial solutions. One may be worried that it would be difficult to find non-trivial solution of SIB among the sea of trivial fixed points. In fact, Ref.~\cite{abdelaleem2023deep} shows that a variational version of SIB easily solves this problem.

In the limit of $\alpha_X,\alpha_Y\to0$, the exponent in the formal solution blows up. As a result, one obtains a deterministic mapping from uncompressed variables to their compressions:
\begin{align}
p(t_x|x) &= \delta(t_x,\tau_x(x))\\ 
\tau_x (x) &= \textrm{argmax}_{t_x}\left[\ln p(t_x)-\beta D_{\rm KL}(p(t_y|x)||p(t_y|t_x))\right],\\
p(t_y|y) &= \delta(t_y,\tau_y(y))\\
\tau_y (y) &= \textrm{argmax}_{t_y}\left[\ln p(t_y)-\beta D_{\rm KL}(p(t_x|y)||p(t_x|t_y))\right].
\end{align}
This is the Deterministic SIB (DSIB).

\section{Results}

To show that GSIB is more data efficient than two GIBs applied independently to $X$ and to $Y$, we notice that, in practical applications, all of the information and entropy terms in the loss functions must be estimated from data. Estimation of information-theoretic quantities is a hard task, potentially as hard as estimating the underlying distributions themselves, largely due to the estimation bias \citep{antos2001convergence, paninski2003estimation}. Crucially, for a DR algorithm to produce meaningful results, the empirically estimated loss function must accurately represent the true loss function, which is unknown to us. Thus the question of which algorithm is more data efficient is equivalent to a different question: for which of the considered IB algorithms does the estimate of the respective loss function converge faster to its true value as the sample size grows?

A lot of ink has been expended on the problem of mutual information estimation \citep{roulston1999estimating,kraskov2004estimating,goebel2005approximation,Hjelm2018}. Here we do not try to produce better estimation techniques. Instead we focus on discrete random variables with finite cardinalities, and we use the simplest estimator,  known as plug-in, naive, or maximum likelihood estimator, for estimation of all of the terms in the loss functions \citep{roulston1999estimating, paninski2003estimation}. For this estimator, which we denote with $\hat{\cdot}$, the probability distribution $p(x)$ is estimated by its maximum likelihood (ML) value, namely  the frequency of an outcome in the sample, $\hat{p}(x)=n(x)/N$, where $n(x)$ is the number of times $x$ occurred, and $N$ is the total number of samples. Then $\hat{H}$, $\hat{I}$, and $\hat{L}$ are all given by plugging in $\hat{p}$ instead of $p$ in the expression for these quantities. \cite{shamir2010learning} showed that, while the ML estimator of mutual information $\hat{I}(X,Y)$ is guaranteed to converge to the true value only when $N\gg |X| |Y|$, the ML estimator of the loss function, $\hat{L}_{\rm IB}$, converges at much smaller $N$, making IB more practical than one would naively think.

Here we continue this line of analysis and examine the convergence properties of $\hat{L}_{\rm GSIB}$ and $\hat{L}_{\rm GIB}$ when both $|X|,|Y|\gg 1$ in two different ways. First, we extend the derivations of \cite{shamir2010learning} and bound the error of estimating each information-theoretic term in each of the loss functions from data. This allows us to build bounds on how close $L$ and $\hat{L}$ are, and we can compare these bounds for GSIB and GIBs. Second, inspired by \cite{still2004many}, we calculate the standard deviation and bias of $L-\hat{L}$ for different versions of the IB. By both measures, for $|X|,|Y|\gg 1$, $\hat{L}_{\rm GSIB}$  will have a smaller bias then $\hat{L}_{\rm GIB}$. This is our main result, allowing us to claim that the symmetric version of IB is more data efficient.

\subsection{Bounds on The Loss Functions}
The loss functions $L_{\rm GSIB}$ and $L_{\rm GIB}$ consist of multiple mutual information and entropy terms. We calculate bounds on the fluctuations between each of these terms and their estimators, and then combine them into a single estimate of the fluctuations of each loss function. We do this below in detail for $I(T_X;X)$ and its estimator $\hat{I}(T_X;X)$. Analysis of the other terms is similar. Furthermore, for our analysis, only the distributions of $x$ and $y$ are unknown, and must be sampled from data. The distributions $p(t_x|x)$ and $p(t_y|y)$ are chosen by the algorithm and optimized over. That is, they are {\em known} in any particular iteration of the scheme. Thus they do not produce fluctuations in the loss function directly, but only through the induced $p(t_x,t_y)$, which fluctuate. This means that, as first noticed in Ref.~\cite{shamir2010learning}, some terms do not contribute to the fluctuation bounds, simplifying the results. Crucially, our expressions below will hold for all mappings $p(t_x|x)$ and $p(t_y|y)$, and not just the mappings that minimize their respective loss functions.

To estimate $|I(T_X;X)-\hat{I}(T_X;X)|$, we compare both terms to the expected value of the empirical information $E(I(T_X;X))$:
\begin{multline}
  |\hat{I}(T_X;X)-{I}(T_X;X)|
  =|\hat{I}(T_X;X)-E(\hat{I}(T_X;X))+E(\hat{I}(T_X;X))-{I}(T_X;X)|\\
\le|\hat{I}(T_X;X)-E(\hat{I}(T_X;X))| +  |I(T_X;X)-E(\hat{I}(T_X;X))|.\label{twoparts}
\end{multline}
This is analogous to the usual bias-variance decomposition for bounds on the magnitude of fluctuations, with the first term in Eq.~(\ref{twoparts}) representing the absolute deviation of the estimator, and the second the bias. We now bound the absolute deviation and the the bias terms separately.

First we focus on the absolute deviation (first) term in  Eq.~(\ref{twoparts}). For this, we follow \cite{shamir2010learning} and rely on the the  McDiarmid's inequality. This concentration inequality bounds the probability of the difference between a function of an empirical sample and its expected value. The bound is constructed from bounds on the change in the function due to changes in individual data points:
\begin{align}
    P\left[\left|f(x_1,x_2,\dots,x_N)- E\left(f(x_1,x_2,\dots,x_N)\right)\right|\geq \epsilon\right]&\le2 \exp\left[-\frac{2\epsilon^2}{\sum c_i}\right]\equiv \delta_1,\label{McDiarmid1}\\
    \mbox{where}\quad\left|f(x_1,\dots,x_i,\dots,x_N)- f(x_1,\dots,x_i',\dots,x_N) \right|&\leq c_i.
    \label{McDiarmid2}
\end{align}
Thus, to use the inequality, we consider the maximum change in $\hat{I}$ if a single datum is changed. That is, suppose the data point $(x,y)$ is replaced by another data point $(x',y')$. Then the maximum likelihood estimator at the point $(x,y)$, $\hat{p}(x,y)$, decreases by $1/N$. In contrast, $\hat{p}(x',y')$ increases by $1/N$, and the estimate does not change at all other $x$, $y$ values. Similarly, the marginals $\hat{p}(x)$, $\hat{p}(x')$, $\hat{p}(y)$, and $\hat{p}(y')$ change by at most $1/N$, while marginals at all other values remain the same. For a fixed compression mapping, we calculate $\hat{p}(t_x)=\sum_x p(t_x|x) \hat{p}(x)$. We see that, with a single datum moving, $\hat{p}(t_x)$ can change by at most $|p((t_x|x')-p(t_x|x))|/N\le1/N$ for each $t_x \in T_X$. Similarly $\hat{p}(t_y)$ can change by at most $1/N$ for each $t_y \in T_Y$. 

We now express the relevant mutual information in terms of entropy, $\hat{I}_{\alpha_X}(T_X;X)=\hat{H}(T_X)-\alpha_X \hat{H}(T_X|X)$,  where the entropy $\hat{H}(T_X)$ depends on the probability density $\hat{p}(t_x)$: 
\begin{equation}\hat{H}(T_X)=-\sum_{t_x} \hat{p}(t_x) \log \hat{p}(t_x).\label{Hhat}\end{equation}
The change in entropy from moving a single datum can be bounded using the following inequality, again borrowed from \cite{shamir2010learning}:
\begin{equation}|(a+\delta)\log(a+\delta)-a \log a|\le \log(N)/N \end{equation}
for any positive integer $N$ and for any $a \in [0,1-1/N]$ and $\delta\le1/N$. We apply this identity for each term in the sum in Eq.~(\ref{Hhat})  and find that the change in $\hat{H}(T_X)$ is bounded by $|T_X|\log N/N$.

We bound the change in $\hat{H}(T_X|X)=\sum_x \hat{p}(x) H(T_X|X=x)$. $H(T_X|X=x)$ only depends on $p(t_x|x)$, which we consider fixed. $\hat{p}(x)$ changes by at most $1/N$ for two values of $x$. Thus the largest change is  \sloppy$|H(T_X|x')-H(T_X|x)|/N\le |\max(H(T_X|x'),H(T_X|x))|/N\le \log |T_X| / N$. The last inequality comes from  \sloppy$H(T_X|X=x)\le \log |T_X|$, with the bound achieved for the uniform distribution. 

Finally, combining the results for both entropy terms,  we see that $\hat{I}_{\alpha_X}(T_X;X)$ can change by at most $(|T_X| \log N + \alpha_X \log |T_X|)/N$. Now we apply the McDiarmid inequality, Eqs.~(\ref{McDiarmid1}, \ref{McDiarmid2}) to finally obtain that, with probability of at least $1-\delta_1$:
\begin{equation}
  |\hat{I}_{\alpha_X}(T_X;X)-E(\hat{I}_{\alpha_X}(T_X;X))|\le (|T_X| \log N + \alpha_X \log |T_X|) \frac{\sqrt{\log(2/\delta_1)}}{\sqrt{2N}}.\end{equation}
This generalizes the result of \cite{shamir2010learning} to $\alpha_X \neq 1$. Similarly, we get that,  with probability of at least $1-\delta_1$, 
\begin{equation}|\hat{I}_{\alpha_Y}(T_Y;Y)-E(\hat{I}_{\alpha_Y}(T_Y;Y))|\le (|T_Y| \log N + \alpha_Y \log |T_Y|) \frac{\sqrt{\log(2/\delta_1)}}{\sqrt{2N}}.\end{equation}

This leaves us with the final bound on the difference between the ML estimators of various informations and their expectations, namely for $\hat{I}(T_X;T_Y)$;  this quantity is not analysed in \cite{shamir2010learning}, but we proceed very similarly. First, we calculate how much this term changes from a single datum being moved by using the identity $\hat{I}(T_X;T_X)=\hat{H}(T_X)+\hat{H}(T_Y)-\hat{H}(T_X,T_Y)$. Luckily we already calculated that  $\hat{H}(T_X)$ changes by, at most, $|T_X|\log N/N$, and $\hat{H}(T_Y)$ changes by, at most, $|T_Y|\log N/N$. We are left to calculate how much $\hat{H}(T_X,T_Y)$ can change.  We write $\hat{H}(T_X,T_Y)=-\sum_{t_x,t_y}\hat{p}(t_x,t_y) \log \hat{p}(t_x,t_y)$, where $\hat{p}(t_x,t_y)=\sum_{x,y}p(t_x|x)p(t_y|y)\hat{p}(x,y)$. Therefore, $\hat{p}(t_x,t_y)$ can change by, at most, $1/N$ for all $(t_x,t_y)\in (T_X,T_Y)$. Thus, $\hat{H}(T_X,T_Y)$ can change by at most $|T_X||T_Y|\log N/N$. We again use the McDiarmid's inequality and we determine that,  with probability of at least $1-\delta_1$, the difference between the ML estimate $\hat{I}(T_X;T_Y)$ and its expected value is bounded by
\begin{equation}|\hat{I}(T_X;T_Y)-E(\hat{I}(T_X;T_Y))|\le ((|T_X| + |T_Y| +|T_X||T_Y|)\log N)\frac{\sqrt{\log(2/\delta_1)}}{\sqrt{2N}}.\end{equation}

Now we need to calculate bounds on the bias (second) terms in Eq.~(\ref{twoparts}) and similar expressions for the other information quantities.  For this, we use results from \cite{paninski2003estimation}, namely:
\begin{align}
|H(T_X)-E(\hat{H}(T_X))|&\le\log\left(1+\frac{|T_X|-1}{N}\right)\le \frac{|T_X|-1}{N},\\
|H(T_Y)-E(\hat{H}(T_Y))|&\le\log\left(1+\frac{|T_Y|-1}{N}\right)\le \frac{|T_Y|-1}{N},\\
|H(T_X,T_X)-E(\hat{H}(T_X,T_X))|&\le\log\left(1+\frac{|T_X||T_Y|-1}{N}\right)\le \frac{|T_X||T_Y|-1}{N}.
\end{align}
Since we consider  mapping $p(t_x|x)$ as fixed and known for this analysis, there is no bias $H(T_X|X)-E(\hat{H}(T_X|X))$. This means that the bias $|I_{\alpha_X}(T_X;X)-\hat{I}_{\alpha_X}(T_X;X)|$ only comes from the $|H(T_X)-\hat{H}(T_X)|$ term and does not have an $|X|$ or $\alpha_x$ dependence.

Putting the bounds on deviations of the estimates from their expectations and of expectations from the true values together, we get bounds on fluctuations of various information quantities that contribute to the  GSIB loss function
\begin{align}
|I_{\alpha_X}(T_X;X)-\hat{I}_{\alpha_X}(T_X;X)|\le& (|T_X| \log N  + \alpha_X \log |T_X|) \frac{\sqrt{\log(2/\delta_1)}}{\sqrt{2N}}+\frac{|T_X|-1}{N},\label{LGSIB1}\\
|I_{\alpha_Y}(T_Y;Y)-\hat{I}_{\alpha_Y}(T_Y;Y)|\le& (|T_Y| \log N + \alpha_Y \log |T_Y|) \frac{\sqrt{\log(2/\delta_1)}}{\sqrt{2N}}+\frac{|T_Y|-1}{N},\label{LGSIB2}\\
|I(T_X;T_Y)-\hat{I}(T_X;T_Y)|\le& (|T_X|+|T_Y|+|T_X||T_Y|)\log N \frac{\sqrt{\log(2/\delta_1)}}{\sqrt{2N}}\nonumber\\&+ \frac{|T_X|-1}{N}+\frac{|T_Y|-1}{N}+\frac{|T_X||T_Y|-1}{N}\nonumber\\
=& ((|T_X|+1)(|T_Y|+1)-1)\log N \frac{\sqrt{\log(2/\delta_1)}}{\sqrt{2N}}\nonumber\\
&+\frac{(|T_X|+1)(|T_Y|+1)-4}{N}\label{LGSIB3}.
\end{align}
For comparison, the term $|I_{\alpha_x}(T_X;X)-\hat{I}_{\alpha_x}(T_X;X)|$ in the error of the GIB loss function has the same bounds as the corresponding term in GSIB, Eq.~(\ref{LGSIB1}). Further the term $|I(T_X;Y)-\hat{I}(T_X;Y)|$ in the error of the GIB loss function is the same as for the traditional IB. \cite{shamir2010learning} calculated it to be:
\begin{align}
|I(T_X;Y)-\hat{I}(T_X;Y)|&\le (3|T_X| + 2) \log N \frac{\sqrt{\log(2/\delta_1)}}{\sqrt{2N}}+\frac{(|T_X|+1)(|Y|+1)-4}{N}\label{LGIB2}.
\end{align}

All of these bounds have a similar structure. The term proportional to $1/\sqrt{N}$ comes from the absolute deviation of the estimators. Its contribution is controlled by $\delta_1$, so that if a high certainty is required ($\delta_1\to 0$), then these terms are large. The terms proportional to $1/N$ are the bias terms. 

The most crucial observation is that, even though the data comes from the joint probability distribution $p(x,y)$, which has the cardinality of $|X||Y|$, the terms proportional to this joint cardinality do not appear in the bounds, 
similar to \cite{shamir2010learning}. In other words, one does not need to have the joint distribution well-sampled to apply any of the IB variants.

The second observation from the bounds is that the deterministic  versions, $\alpha=\alpha_X=\alpha_Y=0$, of both the SIB and the IB have slightly tighter bounds than their generalized counterparts, including the original IB versions with $\alpha=\alpha_X=\alpha_Y=1$. The tightening does not affect the bias component of the bounds, but provides a small correction to the absolute deviation, eliminating the terms similar to $\alpha \log|T_X| \frac{\sqrt{\log(2/\delta_1)}}{\sqrt{2N}}$, which are subdominant in the size of the reduced representations compared to the terms like $|T_X| \log N \frac{\sqrt{\log(2/\delta_1)}}{\sqrt{2N}}$. 

We now compare the data efficiency of GSIB with that of two GIBs applied to reduce $X$ and $Y$ independently. We do so by bounding the error of the estimates of the loss for the GSIB vs.\ for  two GIBs run in parallel. 

The GSIB loss function error is:
\begin{multline}
|L_{\rm GSIB}-\hat{L}_{\rm GSIB}|\le \left(\left(|T_X|+|T_Y|\right) \log N  + \alpha_X \log |T_X| + \alpha_Y \log |T_Y|\right) \frac{\sqrt{\log(2/\delta_1)}}{\sqrt{2N}}\\
+ \beta \left(\left(|T_X|+1\right)\left(|T_Y|+1\right)-1\right)\log N \frac{\sqrt{\log(2/\delta_1)}}{\sqrt{2N}}\\
+\frac{|T_X|-1}{N}+\frac{|T_Y|-1}{N}+\beta \frac{(|T_X|+1)(|T_Y|+1)-4}{N}\label{LGSIBEQ}.
\end{multline}

The combined loss of two GIBs reducing $X$ and $Y$ independently is:
\begin{multline}
|L_{\rm GIB}-\hat{L}_{\rm GIB}|\le \left(\left(|T_X|+|T_Y|\right) \log N  + \alpha_X \log |T_X| + \alpha_Y \log |T_Y|\right) \frac{\sqrt{\log(2/\delta_1)}}{\sqrt{2N}}\\
+ \beta \left(3|T_X|+3|T_Y| + 4\right)\log N \frac{\sqrt{\log(2/\delta_1)}}{\sqrt{2N}}\\
+\frac{|T_X|-1}{N}+\frac{|T_Y|-1}{N}+\beta \frac{(|T_X|+1)(|Y|+1)+(|T_Y|+1)(|X|+1)-8}{N}\label{LGIBEQ}.
\end{multline}
We see that the dominant contribution to the absolute deviation part of $L_{\rm GSIB}$ bound is  $\beta|T_X||T_Y| \log N\frac{\sqrt{\log(2/\delta_1)}}{\sqrt{2N}}$. For two GIBs run in parallel, Eq.~(\ref{LGIBEQ}) says that the dominant contributions to the absolute deviation would be $3\beta(|T_X|+|T_Y|)\log N \frac{\sqrt{\log(2/\delta_1)}}{\sqrt{2N}}$. That is, the two GIBs have smaller absolute deviations than GSIB for all but the smallest cardinalities of the compressed variables. However, notice that the cardinality of the compressed variables is usually not large, almost by definition, so that this loosening of the bound may be too small to notice for realistic $N\gg1$. The behavior of the bias contributions to the bounds is different. The leading term for GSIB is $|T_X||T_Y|/N$, while for two GIBs it is $(|T_X||Y|+|X||T_Y|)/N$. Thus, when $|X|,|Y|\sim N$,  the GSIB can be {\em significantly} more efficient that GIBs. When $|X|,|Y|\gg N$, the bias bounds for GIBs become meaningless, but GSIB bounds do not depend on the cardinality of the data variables. This is the reason for our assertion that GSIB has better data efficiency than two GIBs run in parallel for realistic cardinalities of variables and sample sizes.

\subsection{Mean error and Mean squared error}
\label{meanError}
The error bounds for the mutual information estimators must hold for worst case underlying distributions. Thus there are many cases when the error is significantly smaller than the calculated bounds. To explore if typical errors are different from the worst case bounds, here we calculate the mean squared error of $L_{\rm GSIB}-\hat{L}_{\rm GSIB}$, and similarly for the  GIB. As always, the mean squared error is the sum of the squared bias and the variance of the estimator
\begin{equation}
E(L_{\rm GSIB}-\hat{L}_{\rm GSIB})^2 = (L_{\rm GSIB}-E(\hat{L}_{\rm GSIB}))^2 + E((\hat{L}_{\rm GSIB}-E(\hat{L}_{\rm GSIB}))^2),
\label{BiasVarMSE}
\end{equation}
and similarly for the GIB. This expression is the bias-variance decomposition and is similar to the bias absolute deviation decomposition for the bounds, Eq.~(\ref{twoparts}). However, instead of bounding terms, we now calculate them.  For this, we  decompose every mutual information term in the loss functions into the corresponding entropy components. 

We use the notation $\delta h \equiv \hat{h} - h$ for any variable that is being estimated via the ML estimator. For the ML estimator of the probability distribution $p(x,y)$, multinomial counting statistics textbook results give 
\begin{align}
E(\delta p(x,y))&=0,\label{edpxy}\\
E(\delta p(x,y)\delta p(x',y'))&=\frac{p(x,y)\delta_{x,x'}\delta_{y,y'}}{N}-\frac{p(x,y)p(x',y')}{N}.\label{edpdpxy}
\end{align}
Expectations for fluctuations of marginal distributions can be obtained by marginalizing Eqs.~(\ref{edpxy},~\ref{edpdpxy}). 

In what follows, we will focus on $N\gg1$, so that fluctuations $\delta p(x,y)$ have a small relative variance. Then, to obtain expressions for the variance of entropies, we follow \cite{still2004many} and expand $\hat{H}$ around the true value $H$ for small $\delta p$. For $H(X)$, we get (expressions for other entropy terms are similar):
\begin{align}
\hat{H}(X)&=-\sum_{X}(p(x)+\delta p(x))\log(p(x)+\delta p(x))\nonumber\\
&=-\sum_{X}\left[p(x)\log p(x)+(\log p(x)+1)\delta p(x)+\sum_{n=2}^{\infty}\frac{(-1)^n(\delta p(x))^n}{n(n-1)p(x)^{n-1}}\right]\nonumber\\
&=H(X)-\sum_{X}\left[(\log p(x)+1)\delta p(x)+\sum_{n=2}^{\infty}\frac{(-1)^n(\delta p(x))^n}{n(n-1)p(x)^{n-1}}\right]\label{deltaH}.
\end{align}

From this, it follows that $\delta H(X) = -\sum_{X}\left[(\log p(x)+1)\delta p(x)+\frac{(\delta p(x))^2}{2 p(x)}\right.$ $\left. + O((\delta p(x))^3))\right]$. Noticing that terms first order in $\delta p$ vanish under averaging with respect to $\delta p$, cf.~Eq.~(\ref{edpxy}), we immediately calculate $|E(\delta H(X))|=\frac{|X|-1}{2N}$ and $|E(\delta H(Y))|=\frac{|Y|-1}{2N}$. Similarly, because $p(t_x|x)$ is fixed, we get  $|E(\delta H(X,T_X))|=\frac{|X|-1}{2N}$, $|E(\delta H(Y,T_Y))|=\frac{|Y|-1}{2N}$. Further,   $|E(\delta H(T_X))|=\frac{\sum_{T_X,X}p(t_x|x)p(x|t_x)- 1}{2N}\le\frac{|T_X|-1}{2N}$ and $|E(\delta H(T_Y))|=\le\frac{|T_Y|-1}{2N}$ where the inequalities comes from $p(t_x|x), p(t_y|y) \le 1 $. Combining these and similar results, we get biases of estimators of mutual information terms, which enter the GSIB loss functions:
\begin{align}
    |E(\delta I_{\alpha_X}(X,T_X))|&\le\frac{|T_X|-1}{2N},\\
    |E(\delta I_{\alpha_Y}(Y,T_Y))|&\le\frac{|T_Y|-1}{2N},\\
    |E(\delta I(T_X,T_Y))|&\le \frac{(|T_X|+1)(|T_Y|+1)-4}{2N}.
\end{align}
For the terms in the GIB loss function, we similarly get
\begin{align}
    |E(\delta I(Y,T_X))|&\le\frac{(|Y|+1)(|T_X|+1)-4}{2N},\\
    |E(\delta I(Y,T_Y))|&\le\frac{(|X|+1)(|T_Y|+1)-4}{2N}.
\end{align}
Note that these biases, to the two leading orders in $\delta p$, are half of  the bound on the biases  obtained in the previous Section, Eqs.~(\ref{LGSIB1}-\ref{LGIB2}). Thus the same scaling analyses apply. Crucially, we again observe that the bias of the symmetric variant of GIB only depends on the cardinalities of the compressed variables and not the uncompressed ones.   Hence it is much smaller than for two GIBs applied in parallel, where the bias depends on $|X||T_Y|$ and $|Y||T_X|$.

Similarly we now calculate the mean squared error (see Appendix for details):
\begin{multline}
E(\delta I(X,T_X)^2)=\\=\frac{1}{N}\left[\sum_{X,T_X,T_X'} p(t_x|x)p(t_x'|x)p(x)\log\frac{p(x,t_x)}{p(x)p(t_x)}\log\frac{p(x,t_x')}{p(x)p(t_x')}-I(X,T_X)^2\right].
\end{multline}
This expression can be simplified in two important limits. First, we consider the trivial minimum of the loss function, discussed earlier. There the mapping is uniform, $p(t_x|x)=1/|T_X|$, so that also $p(t_x)=1/|T_X|$. We get:
\begin{multline}
E(I(X,T_X)-\hat{I}(X,T_X))^2=\\= \sum_{X,T_X,T_X'}\frac{p(x)}{N|T_X|^2}\log\frac{p(x)/|T_X|}{p(x)/|T_X|}\log\frac{p(x)/|T_X|}{p(x)/|T_X|}-\frac{0^2}{N}
=0.
\end{multline}
That is, fluctuations vanish in this case. This is expected since there is no information between $T_X$ and $X$, and measuring more data points does  not result in a more accurate estimate of the mutual information.

The second interesting case is a ``winner-take-all'' mapping,  $p(t_x|x)=\delta(t_x,\tau(x))$, which would correspond to a deterministic clustering of multiple values of $x$ into one $t_x$. This results in 
\begin{align}
E(I(X,T_X)-\hat{I}(X,T_X))^2&=\frac{1}{N}\left[\sum_{X} p(x)\log\frac{1}{p(\tau(x))}\log\frac{1}{p(\tau(x))}-I(X,T_X)^2\right]\nonumber\\
&\le \frac{1}{N}\left[\log(\min(|T_X|,|X|))^2 -I(X,T_X)^2\right].
\end{align}
Thus, here the average squared error is bound by $\frac{\log|T_X|^2-I(X,T_X)^2}{N}\le\frac{\log|T_X|^2}{N}$, which means that the RMS error for $I(T_X,X)$ is  $\le\frac{\log|T_X|}{\sqrt{N}}$. Similarly, the RMS errors for $I (T_Y,Y)$ and $I(T_X,T_Y)$ are  $\le\frac{\log|T_Y|}{\sqrt{N}}$ and  $\le\frac{\log\min(|T_X|,|T_Y|)}{\sqrt{N}}$, respectively. For the traditional IB, the RMS error for $I(T,X)$ is $\le\frac{\log|T|}{\sqrt{N}}$, and the RMS error for $I(T,Y)$ is  $\le\frac{\log|T|}{\sqrt{N}}$. Thus, the average fluctuations are  small and are of the same order of magnitude for both the symmetric bottleneck and the traditional bottleneck. This means that the dominant term is the average bias. As we saw earlier, the latter can be much worse for the traditional IB than for the symmetric IB.

\section{Conclusion}
Here we defined the generalized symmetric version of the information bottleneck (GSIB). We  calculated the error bounds for each term within the loss function of GSIB and of the loss functions of the traditional generalized information bottleneck (GIB). We showed that the bias in estimating the loss function, and hence the error in finding the solution to the optimization problem from a finite dataset, is smaller for the GSIB compared to applying traditional GIB to each of the input variables, in parallel. We also calculated the average error and RMS error for each of these terms, resulting in essentially the same conclusions. All of these results suggest that when the cardinality of the measured variables $X$ and $Y$ are both large, and both variables require compression, then simultaneous compression is more data efficient than independently compressing each of the input variables. 

While making extrapolations from a simple discrete variable case to more complex scenarios is difficult, we hope that these results are only the first of many to demonstrate a more general point that {\em simultaneous} dimensionality reduction is typically more data efficient than {\em independent} dimensionality reduction. In fact, using numerical simulations, we recently demonstrated a very similar result for a class of linear dimensionality reduction techniques for continuous variables \cite{abdelaleem2023simultaneous}, as well as for variational autoencoders (an IDR method) and a variational version of SIB (an SDR method) for large-dimensional continuous variables \cite{abdelaleem2023deep}. Collectively, these  findings suggest a general paradigm for efficient dimensionality reduction in complex multivariate datasets. For example, since physical theories are often formulated in terms of collective, coarse-grained representations (e.g., magnetization or temperature, which are expectation values of microscopic spins or energies of molecules), existence of data efficient algorithms for finding such reduced representations bodes well for using data-driven approaches for building physical theories of complex systems. Similarly, in biology, many central questions can be formulated as finding relations between large dimensional datasets. For example, in neuroscience, one aims to relate neural activity to behavior   \citep{Harris2021, Churchland2022, Poeppel2017, Fairhall2015}, and in systems biology, one looks to relate the gene expression state of a cell to its phenotypic profile  \citep{Prior2013, Bielas2017, Teichmann2018, O'Donovan2015, forthe2018}.  Our analysis suggests that methods based on the simultaneous dimensionality reduction can have a substantial impact on these fields as well. 

\subsection*{Acknowledgments}
The authors are grateful to Eslam Abdelaleem and Ahmed Roman for useful discussions and Sean Ridout  for providing feedback on the manuscript. This work was supported, in part, by the Simons Investigator award and NSF Grant No. 2010524.

\section{Appendix}
\subsection{Appendix: Derivation of the Generalized Symmetric Bottleneck}

In what follows, we will derive the formal solution for the generalized symmetric bottleneck for $p(t_x|x)$. The formal solution is found by minimizing the cost function, Eq.~(\ref{eq:gsib}) with respect to $p(t_x|x)$, subject to the normalization constraint. For this, we calculate the following useful derivatives:
\begin{align}
\frac{\partial p(t_x)}{\partial p(t_x'|x')}&=\frac{\partial}{\partial p(t_x'|x')}\sum_X p(t_x|x) p(x)=\delta(t_x,t_x')p(x'),\\
\frac{\partial p(t_y)}{\partial p(t_x'|x')}&=0,\\
\frac{\partial p(t_x,t_y)}{\partial p(t_x'|x')}&=\frac{\partial}{\partial p(t_x'|x')}\sum_{X} p(t_x|x) p(x,t_y)=\delta(t_x,t_x')p(x',t_y).
\end{align}

To enforce the normalization of $p(t_x|x)$, we add a Lagrange multiplier $\lambda$ times  the normalization constraint to the cost function. With the helpful identities above, we now find the first  derivative:
\begin{align}
&\frac{\partial (L_{\rm GSIB}+\lambda (\sum_{X,T_X} p(t_x|x)p(x) - 1))}{\partial p(t_x'|x')} \nonumber=\\
&\quad\quad\quad=\frac{\partial}{\partial p(t_x'|x')}\left[-\sum_{T_X}p(t_x)\ln p(t_x)+\alpha_x \sum_{X,T_X}p(x)p(t_x|x)\ln p(t_x|x)\right.\nonumber\\ 
&\quad\quad\quad\quad\quad\left.-\sum_{T_Y}p(t_y)\ln p(t_y)+\alpha_y \sum_{Y,T_Y}p(y)p(t_y|y)\ln p(t_y|y)\right.\nonumber\\
&\quad\quad\quad\quad\quad\left. -\beta \sum_{T_X,T_Y} p(t_x,t_y)\ln\frac{p(t_x,t_y)}{p(t_x)p(t_y)}+\lambda \left(\sum_{X,T_X} p(t_x|x)p(x) - 1\right)\right] \nonumber\\
&\quad\quad\quad=-p(x')\ln p(t_x') -p(x') +\alpha_x [p(x')\ln p(t_x'|x')+p(x')] \nonumber\\
&\quad\quad\quad\quad\quad- \beta \sum_{T_Y} p(x',t_y) \ln\frac{p(t_x',t_y)}{p(t_x')p(t_y)} + \lambda p(x')\nonumber\\
&\quad\quad\quad=-p(x')\left[\ln p(t_x')+1-\lambda -\alpha_x \left(\ln p(t_x'|x')+1\right)\right.\nonumber\\
&\left.\quad\quad\quad\quad\quad+ \beta \sum_{T_Y} p(t_y|x') \ln \frac{p(t_y|t_x')}{p(t_y)}\frac{p(t_y|x')}{p(t_y|x')}\right]\nonumber\\
&\quad\quad\quad=-p(x')\left[\ln p(t_x'))+1-\lambda -\alpha_x \left(\ln p(t_x'|x')+1\right)\right.\nonumber\\
&\quad\quad\quad\quad\left.+ \beta \sum_{T_Y} p(t_y|x')\ln \frac{p(t_y|t_x')p(t_y|x')}{p(t_y)p(t_y|x')}\right]\nonumber\\
&\quad\quad\quad=-p(x')\left[\ln p(t_x') +1-\lambda -\alpha_x \left(\ln p(t_x'|x')+1\right)\right.\nonumber\\
&\quad\quad\quad\quad\quad\left.+ \beta \sum_{T_Y} p(t_y|x') \left(\ln \frac{p(t_y|x')}{p(t_y)}-\ln\frac{p(t_y|x')}{p(t_y|t_x')}\right)\right]\nonumber\\
&\quad\quad\quad=-p(x')\left[\ln p(t_x') +1-\lambda -\alpha_x \left(\ln p(t_x'|x')+1\right)\right.\nonumber\\ 
&\left.\quad\quad\quad\quad\quad+ \beta D_{\rm KL}(p(t_y|x')||p(t_y))-\beta D_{\rm KL}(p(t_y|x')||p(t_y|t_x'))\right].
\end{align}

We now find the minimum of the cost function subject to the constraint that $p(t_x|x)$ is normalized by setting this  derivative to zero and solving for $p(t_x'|x')$. Doing this, we find a formal solution:
\begin{equation}
p(t_x'|x')=\frac{\exp\left[\frac{1}{\alpha_x}\left(\ln p(t_x') -\beta D_{\rm KL}(p(t_y|x')||p(t_y|t_x')\right)\right]}{Z_x(x',\alpha_x,\beta)},
\end{equation}
where $Z_x(x',\alpha_x,\beta)=\exp\left[-1+\lambda +\alpha_x - \beta D_{\rm KL}(p(t_y|x')||p(t_y))\right]$, and $\lambda$ is chosen such that $p(t_x'|x')$ is normalized. Notice that the normalization constant $Z_x$ is independent of $t_y$ and $t_x'$. It only depends on $x'$, $\alpha_x$, and $\beta$. The same procedure can be followed to find the solution of the generalized symmetric information bottleneck for $p(t_y|y)$.
\begin{equation}
p(t_y'|y')=\frac{\exp\left[\frac{1}{\alpha_y}\left(\ln p(t_y') -\beta D_{\rm KL}(p(t_x|y')||p(t_x|t_y')\right)\right]}{Z_y(y',\alpha_y,\beta)},
\end{equation}

\subsection{Appendix: Mean Error}
Here we make explicit the calculations started in Section~\ref{meanError}. Using Eq.~(\ref{deltaH}) from the main text we, find the expected bias for $X$ to depend on the cardinality $|X|$ and to be:
\begin{align}
|E(\delta H(X))|&=\sum_{X}\frac{E(\delta p(x)^2)}{2 p(x)}
=\sum_{X}\frac{E(\sum_{Y} \delta p(x,y))^2}{2 \sum_{Y} p(x,y)}\nonumber\\
&=\sum_{X}\frac{\sum_{Y,Y'} E(\delta p(x,y)\delta p(x,y'))}{2 \sum_{Y} p(x,y)}\nonumber\\
&=\sum_{X}\frac{\sum_{Y} p(x,y)-\sum_{Y,Y'} p(x,y) p(x,y')}{2N \sum_{Y} p(x,y)}\nonumber\\
&=\sum_{X}\frac{p(x)- p^2(x)}{2N p(x)}=\frac{|X|-1}{2N}.
\end{align}
Similarly, $|E(\delta H(Y))|=\frac{|Y|-1}{2N}$, and $|E(\delta H(X,T_X))|=\frac{|X|-1}{2N}$.

Now we write:
\begin{align}
|E(\delta H(T_X))|&=\sum_{T_X}\frac{E(\delta p(t_x)^2)}{2 p(t_x)}=\sum_{T_X}\frac{E(\sum_{X,Y} \delta p(t_x|x)p(x,y))^2}{2 \sum_{X,Y} p(x,y)}\nonumber\\
&=\sum_{T_X}\frac{\sum_{X,X',Y,Y'} E(p(t_x|x)p(t_x|x')\delta p(x,y)\delta p(x',y'))}{2 \sum_{X,Y} p(t_x|x)p(x,y)}\nonumber\\
&=\sum_{T_X}\frac{\sum_{X,Y} p(t_x|x)^2p(x,y)-\sum_{X,X',Y,Y'} p(t_x|x)p(t_x|x')p(x,y) p(x',y')}{2N \sum_{X,Y} p(t_x|x)p(x,y)}\nonumber\\
&=\sum_{T_X}\frac{\sum_{X}p(t_x|x)^2p(x)- p(t_x)^2}{2Np(t_x)}=\sum_{T_X}\frac{\sum_{X}p(t_x|x)p(x|t_x)- p(t_x)}{2N}\nonumber\\
&=\frac{\sum_{T_X,X}[p(t_x|x)p(x|t_x)]- 1}{2N}\le\frac{|T_X|-1}{2N},
\end{align}
where the inequality comes from $p(t|x) \le 1 $, so that $p(t|x)^2 p(x)\le p(t|x)p(x)$. 

We can combine these results to find the overall bias for $\hat{I}(X,T_X)$:
\begin{align}
|E(\delta I(X,T_X))|&=|E(\delta H(X))+E(\delta H(T_X))-E(\delta H(X,T_X))|\nonumber\\
&=\frac{|X|-1}{2N}+\frac{\sum_{T_X,X}p(t_x|x)p(x|t_x)- 1}{2N}-\frac{|X|-1}{2N}\nonumber\\
&=\frac{\sum_{T_X,X}p(t_x|x)p(x|t_x)- 1}{2N}\le\frac{|T_X|-1}{2N}.
\end{align}
Similarly,
\begin{align}
|E(\delta I(Y,T_Y))|&=|E(\delta H(Y))+E(\delta H(T_Y))-E(\delta H(Y,T_Y))|\nonumber\\
&=\frac{|Y|-1}{2N}+\frac{\sum_{T_Y,Y}p(t_y|y)p(y|t_y)- 1}{2N}-\frac{|Y|-1}{2N}\nonumber\\
&=\frac{\sum_{T_Y,Y}p(t_y|y)p(y|t_y)- 1}{2N}\le\frac{|T_Y|-1}{2N}.
\end{align}

Finally, we  calculate the bias for $\hat{I}(T_X,T_Y)$:
\begin{align}
|E(\delta I(T_X,T_Y))|=&|E(\delta H(T_X))+E(\delta H(T_Y))-E(\delta H(T_X,T_Y))|\nonumber\\
=&\frac{\sum_{T_X,X}p(t_x|x)p(x|t_x)- 1}{2N}+\frac{\sum_{T_Y,Y}p(t_y|y)p(y|t_y)- 1}{2N}\\
&-\frac{\sum_{T_X,T_Y,X,Y}p(t_x,t_y|x,y)p(x,y|t_x,t_y)- 1}{2N},
\end{align}
and
\begin{align}
|E(\delta I(T_X,T_Y))|&\le|E(\delta H(T_X))|+|E(\delta H(T_Y))|+|E(\delta H(T_X,T_Y))|\nonumber\\
&\le\frac{|T_X|-1}{2N}+\frac{|T_Y|-1}{2N}+\frac{|T_X||T_Y|-1}{2N}.
\end{align}

We can perform similar calculations for the original bottleneck to obtain:
\begin{align}
|E(\delta I(Y,T))|&\le |E(\delta H(Y))|+|E(\delta H(T))|+|E(\delta H(Y,T))|\nonumber\\
&=\frac{|Y|-1}{2N}+\frac{\sum_{T,X}p(t|x)p(x|t)- 1}{2N}+\sum_{Y,T}\frac{\sum_{X} p(t|x)p(x|t,y)}{2N}-\frac{1}{2N}\nonumber\\
&\le\frac{|Y|-1}{2N}+ \frac{|T|-1}{2 N} + \frac{|Y||T|-1}{2 N}.
\end{align}

\subsection{Appendix: Mean Squared Error}

Using a method inspired by \cite{still2004many}, we start by calculating the expected squared error for the mutual information between two arbitrary variables $A$ and $B$, where the estimated probabilities are different from the true ones by a small error $\delta$, $\hat{p}(a,b)=p(a,b)+\delta p(a,b)$, $\hat{p}(a)=p(a)+\delta p(a)$ and $\hat{p}(b)=p(b)+\delta p(b)$. First, let's calculate the mutual information to the  first order in $\delta p$:
\begin{align}
\hat{I}(A,B)&=\sum_{A,B}(p(a,b)+\delta p(a,b))\log\frac{p(a,b)+\delta p(a,b)}{(p(a)+\delta p(a))(p(b)+\delta p(b))}\nonumber\\
&=\sum_{A,B}(p(a,b)+\delta p(a,b))\log\left[\frac{p(a,b)}{p(a)p(b)}\frac{1+\delta p(a,b)/p(a,b)}{(1+\delta p(a)/p(a))(1+\delta p(b)/p(b))}\right]\nonumber\\
&=\sum_{A,B}(p(a,b)+\delta p(a,b))\left[\log\frac{p(a,b)}{p(a)p(b)}+\log\left(1+\frac{\delta p(a,b)}{p(a,b)}\right) \right.\nonumber\\
&\left.\quad\quad\quad-\log\left(1+\frac{\delta p(a)}{p(a)}\right) -\log\left(1+\frac{\delta p(b)}{p(b)}\right)\right]\nonumber\\
&\approx \sum_{A,B}(p(a,b)+\delta p(a,b))\left[\log \frac{p(a,b)}{p(a)p(b)}+\frac{\delta p(a,b)}{p(a,b)} -\frac{\delta p(a)}{p(a)}-\frac{\delta p(b)}{p(b)}+\dots\right]\nonumber\\
&\approx  \sum_{A,B}\left[\delta p(a,b)\log \frac{p(a,b)}{p(a)p(b)}+p(a,b)\left(\frac{\delta p(a,b)}{p(a,b)} -\frac{\delta p(a)}{p(a)}-\frac{\delta p(b)}{p(b)}+\dots\right)\right]\nonumber\\
&=\sum_{A,B}\left[\delta p(a,b)\log\frac{p(a,b)}{p(a)p(b)}+\left(\delta p(a,b) -p(b|a)\delta p(a)-p(a|b)\delta p(b))+\dots\right)\right]\nonumber\\
&\quad\quad\quad+ I(A,B) \nonumber\\
&=\sum_{A,B}\delta p(a,b)\log\frac{p(a,b)}{p(a)p(b)} + \sum_{A,B}\delta p(a,b)-\sum_{A}\delta p(a)-\sum_{B}\delta p(b)+\dots\nonumber\\
&\quad\quad\quad+ I(A,B) \nonumber\\
&= I(A,B) +  \sum_{A,B}\delta p(a,b)\left(\log\frac{p(a,b)}{p(a)p(b)}-1\right)+\dots
\end{align}
Where in the last two lines, we used $\sum_B p(b|a)=1$, $\sum_A p(a|b)=1$, and $\delta p(a) = \sum_B \delta p(a,b)$, $\delta p(b) = \sum_A \delta p(a,b)$, respectively.

Thus, we see that $\delta I(A,B) = \sum_{A,B}\delta p(a,b)(\log\frac{p(a,b)}{p(a)p(b)}-1)$ to first order in $\delta p(a,b)$. We can now calculate the average squared error:
\begin{align}
E[\delta I(A,B)^2]&=E\left[\sum_{A,B}\delta p(a,b)\left(\log \frac{p(a,b)}{p(a)p(b)}-1\right)\right.\nonumber\\
&\left.\quad\quad\quad\quad\times\sum_{A',B'}\delta p(a',b')\left(\log\frac{p(a',b')}{p(a')p(b')}-1\right)\right]\nonumber\\
&=\sum_{A,B,A',B'} E\left[\delta p(a,b)\delta p(a',b')\right] \nonumber\\
&\quad\quad\quad\quad\times\left(\log\frac{p(a,b)}{p(a)p(b)}-1\right)\left(\log\frac{p(a',b')}{p(a')p(b')}-1\right).\label{sqInfo}
\end{align}

We can  use this generic expression to find the squared error for the estimator of information between the variables $X$ and $T_X$, where $\delta p(x,t_x) = p(t_x|x)\delta p(x)$, and $E(\delta p(x)\delta p(x'))=1/N[\delta(x,x')p(x)-p(x)p(x')]$. We calculate $E[\delta I(X,T_X)^2]$ as follows: 
\begin{align}
&E[\delta I(X,T_X)^2]\nonumber\\
&=\sum_{X,T,X',T'} E\left[\delta p(x,t_x)\delta p(x',t_x')\right] \left(\log\frac{p(x,t_x)}{p(x)p(t_x)}-1\right)\left(\log\frac{p(x',t_x')}{p(x')p(t_x')}-1\right)\nonumber\\
&=\sum_{X,T_X,X',T_X'} p(t_x|x)p(t_x'|x')E\left[\delta p(x)\delta p(x')\right] \left(\log\frac{p(x,t_x)}{p(x)p(t_x)}-1\right)\nonumber\\
&\quad\quad\quad\quad\times\left(\log\frac{p(x',t_x')}{p(x')p(t_x')}-1\right)\nonumber\\
&=\sum_{X,T_X,X',T_X'} p(t_x|x)p(t_x'|x')\frac{p(x)\delta(x,x')-p(x)p(x')}{N}\nonumber\\
&\quad\quad\quad\quad\times\left(\log\frac{p(x,t_x)}{p(x)p(t_x)}-1\right)\left(\log\frac{p(x',t_x')}{p(x')p(t_x')}-1\right)\nonumber\\
&=\frac{1}{N}\left[\sum_{X,T_X,T_X'} p(t_x|x)p(t_x'|x)p(x)\right.\nonumber\\
&\left.\quad\quad\quad\quad\times\left(\log\frac{p(x,t_x)}{p(x)p(t_x)}\log\frac{p(x,t_x')}{p(x)p(t_x')}-\log\frac{p(x,t_x)}{p(x)p(t_x)}-\log\frac{p(x,t_x')}{p(x)p(t_x')}+1\right)\right]\nonumber\\
&-\frac{1}{N}\left[\sum_{X,T_X} p(t_x|x)p(x)\left(\log\frac{p(x,t_x)}{p(x)p(t_x)}-1\right)\right.\nonumber\\
&\left.\quad\quad\quad\quad\times\sum_{X',T_X'} p(t_x'|x')p(x')\left(\log\frac{p(x',t_x')}{p(x')p(t_x)'}-1\right)\right]\nonumber\\
&=\frac{1}{N}\left[\sum_{X,T_X,T_X'} p(t_x|x)p(t_x'|x)p(x)\log\frac{p(x,t_x)}{p(x)p(t_x)}\log\frac{p(x,t_x')}{p(x)p(t_x')}\right.\nonumber\\
&\left.\quad\quad\quad\quad-2I(X,T_X)+1-(I(X,T_X)-1)^2\right]\nonumber\\
&=\frac{1}{N}\left[\sum_{X,T_X,T_X'} p(t_x|x)p(t_x'|x)p(x)\log\frac{p(x,t_x)}{p(x)p(t_x)}\log\frac{p(x,t_x')}{p(x)p(t_x')}-I(X,T_X)^2\right].\label{EdI2XTx}
\end{align}

Now let's look at two limits when we can simplify the above expression. In the first limit, we  assume that the mapping is uniform, $p(t_x|x)=1/|T_X|$, which means that $p(t_x)=1/|T_X|$ as well. Then
\begin{equation}
E[(I(X,T_X)-\hat{I}(X,T_X))^2]= \sum_{X,T_X,T_X'}\frac{p(x)}{|T_X|^2}\log\frac{p(x)/|T_X|}{p(x)/|T_X|}\log\frac{p(x)/|T_X|}{p(x)/|T_X|}\frac{1}{N}-\frac{0^2}{N}
=0.
\end{equation}
In the  other limit, we assume a ``winner-take-all'' mapping, where $p(t_x|x)=\delta(t_x,\tau(x))$.  We can reduce the expression to:
\begin{align}
&E[\delta I(X,T_X)^2]=\nonumber\\
&\quad\quad=\frac{1}{N}\left[\sum_{X,T_X,T_X'} \delta(t_x,\tau(x))\delta(t_x',\tau(x))p(x)\log\frac{\delta(t_x,\tau(x))}{p(t_x)}\log\frac{\delta(t_x',\tau(x))}{p(t_x')}\right.\nonumber\\
&\quad\quad\quad\quad\quad\left.-I(X,T_X)^2\right]\nonumber\\
&\quad\quad=\frac{1}{N}\left[\sum_{X} p(x)\log\frac{1}{p(\tau(x))}\log\frac{1}{p(\tau(x))}-I(X,T_X)^2\right]\nonumber\\
&\quad\quad\le \frac{1}{N}\left[\log(\min(|T_X|,|X|))^2 -I(X,T_X)^2\right]\le \frac{1}{N}\left[\log(\min(|T_X|,|X|))^2\right].
\end{align}

The result for $E[\delta I(Y,T_Y)^2]$ is  similar to that for  $E[\delta I(X,T_X)^2]$, Eq.~(\ref{EdI2XTx}:
\begin{multline}
E[\delta I(Y,T_Y)^2]=\\=\frac{1}{N}\left[\sum_{Y,T_Y,T_Y'} p(t_y|y)p(t_y'|y)p(y)\log\frac{p(y,t_y)}{p(y)p(t_y)}\log\frac{p(y,t_y')}{p(y)p(t_y')}-I(Y,T_Y)^2\right].
\end{multline}

Finally we can calculate the covariance of fluctuations in the  compressed variables, $T_X$ and $T_Y$. Here $\delta p(t_x,t_y)=\sum_{X,Y} p(t_x|x)p(t_y|y) \delta p(x,y)$, and
\begin{align}
E[\delta p(t_x,t_y) \delta p(t_x',t_y')]&=E\left[\sum_{X,Y} p(t_x|x)p(t_y|y) \delta p(x,y)\sum_{X',Y'} p(t_x'|x')p(t_y'|y') \delta p(x',y')\right]\nonumber\\
&=\sum_{X,Y,X',Y'}  p(t_x|x)p(t_y|y)p(t_x'|x')p(t_y'|y')E[\delta p(x,y) \delta p(x,y)]\nonumber\\
&=\sum_{X,Y,X',Y'}  p(t_x|x)p(t_y|y)p(t_x'|x')p(t_y'|y')\nonumber\\
&\quad\quad\quad\times\frac{p(x,y)\delta(x,x')\delta(y,y')-p(x,y)p(x',y')}{N}\nonumber\\
&=\left[\frac{\sum_{X,Y}  p(t_x|x)p(t_y|y)p(t_x'|x)p(t_y'|y)p(x,y)}{N}\right] \nonumber\\
&\quad\quad\quad-\left[\frac{p(t_x,t_y)p(t_x',t_y')}{N}\right].
\end{align}
Using the previous result and Eq.~(\ref{sqInfo}), we find:
\begin{align}
E[\delta I(T_X,T_Y)^2]&=\sum_{T_X,T_Y,T_X',T_Y'} E[\delta p(t_x,t_y)\delta p(t_x',t_y')] \left(\log\frac{p(t_x,t_y)}{p(t_x)p(t_y)}-1\right)\nonumber\\ &\quad\quad\quad\times\left(\log\frac{p(t_x',t_y')}{p(t_x')p(t_y')}-1\right)\nonumber\\
&=\sum_{T_X,T_Y,T_X',T_Y'}\left[\frac{\sum_{X,Y}  p(t_x|x)p(t_y|y)p(t_x'|x)p(t_y'|y)p(x,y)}{N}\right.\nonumber\\ &\quad\quad\quad\times\left.\log\frac{p(t_x,t_y)}{p(t_x)p(t_y)}\log\frac{p(t_x',t_y')}{p(t_x')p(t_y')} \right]- I(T_X,T_Y)^2/N.
\end{align}

In the ``winner-take-all'' limit, where $p(t_x|x)=\delta(t_x,\tau_x(x))$, and $p(t_y|y)=\delta(t_y,\tau_y(y))$, we find:
\begin{align}
&E[\delta I(T_X,T_Y)^2]=\nonumber\\
&\quad\quad=\sum_{T_X,T_Y,T_X',T_Y'}\left[\frac{\sum_{X,Y}  \delta(t_x,\tau_x(x))\delta(t_y,\tau_y(y))\delta(t_x',\tau_x(x))\delta(t_y',\tau_y(y))p(x,y)}{N}\right.\nonumber\\ &\quad\quad\quad\quad\quad\times\left.\log\frac{p(t_x,t_y)}{p(t_x)p(t_y)}\log\frac{p(t_x',t_y')}{p(t_x')p(t_y')} \right]- I(T_X,T_Y)^2/N\nonumber\\
&\quad\quad=\sum_{X,Y}\left[\frac{p(x,y)}{N}\log\left(\frac{p(\tau_x(x),\tau_y(y))}{p(\tau_x(x))p(\tau_y(y))}\right)^2 \right]- I(T_X,T_Y)^2/N\nonumber\\
&\quad\quad\le \frac{1}{N}\log\left(\min(|T_X|,|T_Y|)\right)^2 - I(T_X,T_Y)^2/N.
\end{align}
Here we have calculate the average bias and variance for each term in the GSIB and the GIB. We found, in general, that the variance decays as $1/N$ and depends only on the cardinality of the compressed variables $|T_X|$ and $|T_Y|$. The expected bias for the GSIB depends on the cardinality of the compressed variables, while the bias for the GIB can depend on both the cardinality of the compressed variables and the cardinality of the uncompressed supervisor variables $|X|$ and $|Y|$.

\bibliographystyle{apalike}
\bibliography{boundsBib2}{}
\end{document}